\documentclass[journal,10pt]{IEEEtran}

\usepackage{amsmath,amssymb,bm,mathtools}
\usepackage{graphicx}
\usepackage{cite}
\usepackage{url}

\newcommand{\bH}{\mathbf H}
\newcommand{\bG}{\mathbf G}
\newcommand{\bJ}{\mathbf J}
\newcommand{\bC}{\mathbf C}
\newcommand{\bD}{\mathbf D}
\newcommand{\bT}{\mathbf T}
\newcommand{\bZ}{\mathbf Z}
\newcommand{\bh}{\mathbf h}
\newcommand{\bg}{\mathbf g}
\newcommand{\bq}{\mathbf q}
\newcommand{\bw}{\mathbf w}
\newcommand{\bx}{\mathbf x}
\newcommand{\bz}{\mathbf z}
\newcommand{\bu}{\mathbf u}
\newcommand{\bv}{\mathbf v}
\newcommand{\be}{\mathbf e}
\newcommand{\cK}{\mathcal K}
\newcommand{\cX}{\mathcal X}
\newcommand{\Tr}{\operatorname{Tr}}
\newcommand{\Rea}{\operatorname{Re}}
\newcommand{\Ima}{\operatorname{Im}}
\newtheorem{lemma}{Lemma}
\newtheorem{proposition}{Proposition}

\newcounter{algorithm}

\title{\vspace{-5pt} \LARGE Move and Countermove: Adversarial Antenna Positioning for Fair Multiuser Downlink}

\author{Lihua Zhu, Guojie Hu, Qingqing Wu, Kui Xu and Zhi Lin%
\thanks{Lihua Zhu and Guojie Hu are with the College of Communication Engineering, Rocket Force
University of Engineering, Xi'an 710025, China (e-mail: hgdxzlh@sina.com; lgdxhgj@sina.com).
Qingqing Wu is with the Department of Electronic Engineering, Shanghai Jiao Tong University, Shanghai 200240,
China (e-mail: qingqingwu@sjtu.edu.cn). Kui Xu is with the College of Communications Engineering, Army Engineering University of PLA, Nanjing 210007, China (e-mail: lgdxxukui@sina.com). Zhi Lin is with the College of Electronic Engineering, National University
of Defense Technology, Hefei 230037, China (e-mail: linzhi945@163.com).}}
\IEEEaftertitletext{\vspace{-20pt}}

\begin{document}
\maketitle
\setlength{\abovedisplayskip}{0.5pt plus 0.5pt}
\setlength{\belowdisplayskip}{0.5pt plus 0.5pt}
\setlength{\abovedisplayshortskip}{0pt plus 0.5pt}
\setlength{\belowdisplayshortskip}{0pt plus 0.5pt}
\setlength{\jot}{0pt}
\setlength{\textfloatsep}{3pt plus 1pt minus 1pt}
\setlength{\floatsep}{3pt plus 1pt minus 1pt}
\setlength{\intextsep}{2pt plus 1pt minus 1pt}
\setlength{\dbltextfloatsep}{3pt plus 1pt minus 1pt}
\setlength{\dblfloatsep}{3pt plus 1pt minus 1pt}
\setlength{\abovecaptionskip}{2pt}
\setlength{\belowcaptionskip}{0pt}

\begin{abstract}
Movable-antenna (MA) technology improves channel conditions through position
adjustment without extra antennas or radio-frequency chains, but the same
reconfigurability can be exploited by a malicious jammer. This paper studies
max--min fair downlink transmission, where a linear-MA base station (BS) serves
multiple single-antenna users against a linear-MA jammer. The BS jointly designs
its antenna positions and user powers, whereas the jammer designs its positions
and jamming beam, yielding a three-stage zero-sum game. For any position pair,
the digital variables are derived in closed form, exactly reducing the game to a nonconvex
position-only min--max problem. We then derive whole-array position gradients
and develop an exchange algorithm: the BS is optimized against a finite attack
set, while separation adds a stronger jammer deployment only when necessary.
Thus, a complete jammer search is avoided at every BS trial point. Numerical
results show a slightly higher independently verified minimum rate than a
nested benchmark with about eight times fewer objective-gradient evaluations.
\end{abstract}


\section{Introduction}
Movable-antenna (MA) technology reshapes wireless channels through local
position adjustment \cite{ZhuOverview2024}. This flexibility can improve
channel discrimination between intended and unintended receivers, motivating
MA-assisted physical-layer security. Existing studies optimize MA positions
for secure MIMO \cite{TangSecure2025}, RIS-assisted sensing and communication
\cite{MaSecureRIS2025}, full-duplex multiuser transmission
\cite{DingSecure2025}, uncertain eavesdropper locations \cite{FengPLS2025},
and confidential/multicast services \cite{ShenPHYSi2025}.

These designs exploit MA position control to improve secrecy against
eavesdroppers. The same control, however, can also be used offensively: a
jammer can reposition its antennas to intensify interference at legitimate
users. Existing work optimizes an MA-enabled jammer against a transmitter
without antenna-position control \cite{HuJammer2025}, whereas conventional
worst-case jamming assumes fixed arrays \cite{GaoWorst2015}. Neither examines
a base station (BS) and jammer that both reposition MAs. BS positioning
changes desired gains and user separability, while jammer positioning
redistributes interference among users. Because these changes affect users
differently, the user with the lowest rate may change as the BS or jammer
adjusts its antenna positions. Consequently, a BS
placement designed for one jammer deployment may lose worst-user protection
after the jammer repositions.

Motivated by this coupled positional competition, we consider a linear-MA BS
serving multiple single-antenna users against a linear-MA jammer. Under
zero-forcing (ZF) beamforming, the BS chooses antenna positions and user powers
to maximize the minimum user rate, while the jammer chooses its positions and
jamming beam to minimize that rate. This leads to a three-stage zero-sum game.
Positions enter the channels through complex exponentials and the ZF gains
through a Gram-matrix inverse. Moreover, one BS placement must withstand a
continuum of jammer deployments. Evaluating each BS candidate therefore
requires a nonconvex worst-case response, resulting in a semi-infinite min--max
problem.

\begin{figure}[t]
\centering
\includegraphics[width=6.3cm]{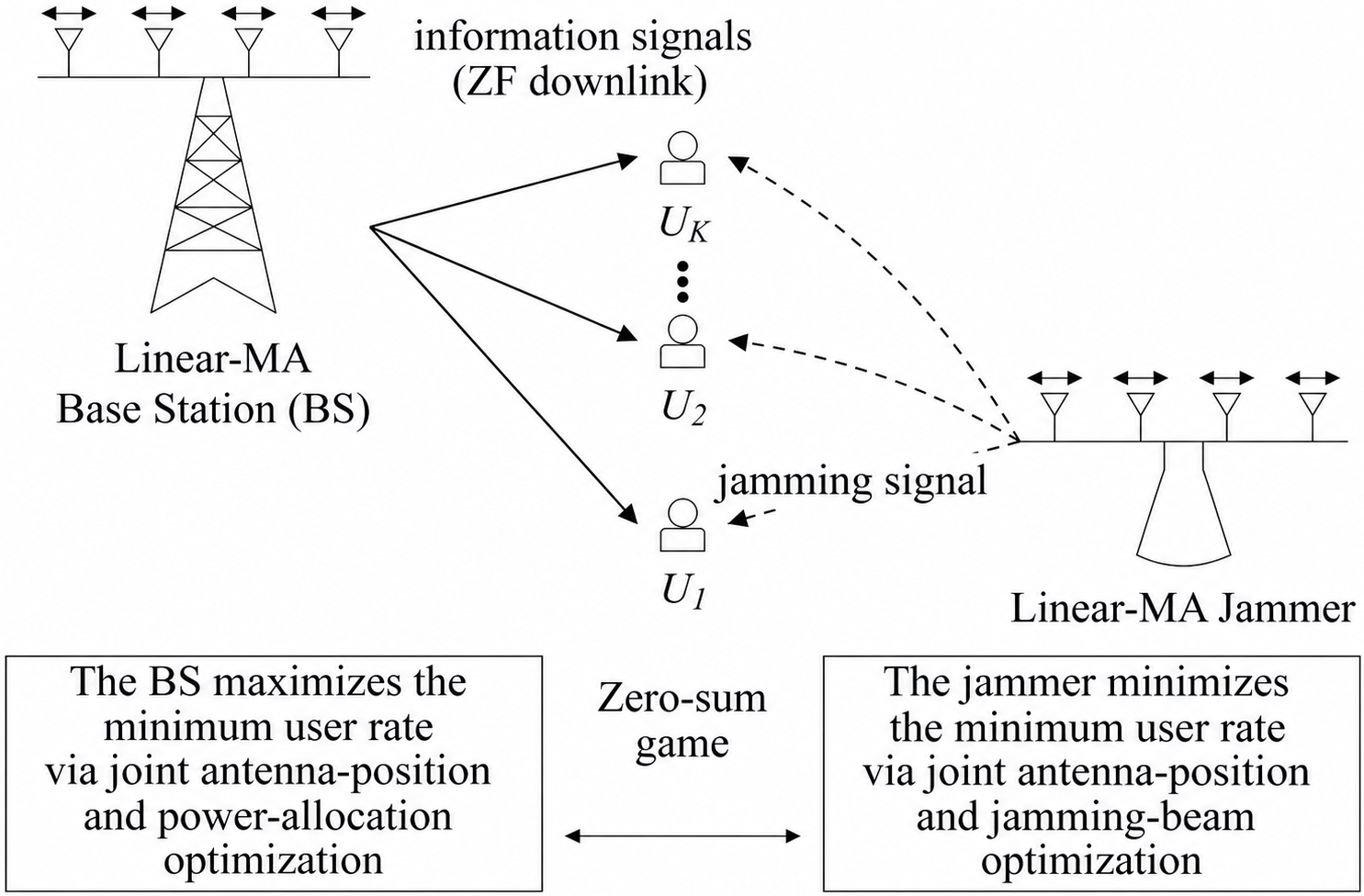}
\caption{System model with a linear-MA BS, $K$ single-antenna users, and a
linear-MA jammer.}
\label{fig:system}
\end{figure}

To address these challenges, the main contributions of this paper are
summarized as follows: i) First, we formulate the sequential zero-sum game;
closed-form ZF power allocation and jamming beam reduce the zero-sum game to
position-only optimization; ii) Building on this reduction, we derive
whole-array gradients by reusing the Gram-matrix inverse and principal
eigenvector; iii) With these gradients, an exchange algorithm alternates finite
BS master and jammer separation steps, with its response gap, projection, and
complexity analyzed; iv) Finally, simulations show a slightly higher
independently verified minimum rate than the nested benchmark
\cite{ZhangExchange2010,ZengCCG2013} with an
$8.3$-fold reduction in objective-gradient evaluations, while outperforming the
FPA-BS/MA-J and Random MA-BS/MA-J baselines.

\section{System Model and Problem Formulation}
\subsection{System Model}

We consider a BS equipped with $N$ transmit MAs serving $K$ single-antenna
users, where $N\ge K$ and $\cK\triangleq\{1,\ldots,K\}$. Meanwhile, a malicious
jammer equipped with $M$ transmit MAs attempts to disrupt the downlink. Let
$x_{q,n}\in\mathbb R$ denote the position of antenna $n$ on the linear
movement region of node $q\in\{{\rm B},{\rm J}\}$, where ${\rm B}$ and
${\rm J}$ denote the BS and jammer, respectively. Define $N_{\rm B}=N$,
$N_{\rm J}=M$, and $\bx_q=[x_{q,1},\ldots,x_{q,N_q}]^T$. The MAs at node
$q$ move over the line segment $[0,L_q]$. Without loss of optimality, the
antennas are indexed according to their spatial order. Hence, the feasible
position set is
\begin{align}
\cX_q\triangleq\{\bx_q:\ &0\le x_{q,1}, x_{q,N_q}\le L_q;\nonumber\\
&x_{q,n+1}-x_{q,n}\ge d_q,\ \forall n<N_q\},             \label{eq:set}
\end{align}
where $d_q>0$ is the minimum antenna spacing and
$L_q\ge(N_q-1)d_q$ ensures nonemptiness.

Since each movement region is much smaller than the link distance, we consider
far-field line-of-sight (LoS) channels. Let $u_k\in[-1,1]$ and
$v_k\in[-1,1]$ denote the direction cosines of the BS--user-$k$ and
jammer--user-$k$ links projected onto their respective array axes. With
$\kappa\triangleq2\pi/\lambda$, where $\lambda$ is the carrier wavelength and
$\mathrm j^2=-1$, the corresponding channels are
\begin{align}
\bh_k(\bx_{\rm B})&=\sqrt{\beta_k}
[e^{\mathrm j\kappa x_{{\rm B},1}u_k},\ldots,
e^{\mathrm j\kappa x_{{\rm B},N}u_k}]^T,                    \label{eq:h}\\
\bg_k(\bx_{\rm J})&=\sqrt{\alpha_k}
[e^{\mathrm j\kappa x_{{\rm J},1}v_k},\ldots,
e^{\mathrm j\kappa x_{{\rm J},M}v_k}]^T,                    \label{eq:g}
\end{align}
where $\beta_k$ and $\alpha_k$ are the channel power gains of the BS--user-$k$
and jammer--user-$k$ links, respectively.

Define $\bH(\bx_{\rm B})=[\bh_1(\bx_{\rm B}),\ldots,
\bh_K(\bx_{\rm B})]$ and
$\bG(\bx_{\rm B})=\bH^H(\bx_{\rm B})\bH(\bx_{\rm B})$.
In this paper, we consider that the BS adopts zero-forcing (ZF) transmission
since: i) it removes
inter-user interference and converts each user's received signal into a
single-user form; ii) it permits the closed-form fair power allocation
derived in Section~III-A, which makes the subsequent game analysis tractable.
The normalized ZF direction and effective channel gain of user $k$ are,
respectively,
\begin{align}
\bw_k(\bx_{\rm B})&=\frac{\bH(\bx_{\rm B})\bG^{-1}(\bx_{\rm B})\be_k}
{\sqrt{[\bG^{-1}(\bx_{\rm B})]_{kk}}},                       \label{eq:zf}\\
a_k(\bx_{\rm B})&=|\bh_k^H(\bx_{\rm B})\bw_k(\bx_{\rm B})|^2
=\frac{1}{[\bG^{-1}(\bx_{\rm B})]_{kk}},                  \label{eq:ak}
\end{align}
where $\be_k$ is the $k$th column of the $K\times K$ identity matrix,
$[\mathbf A]_{kk}$ denotes the $k$th diagonal entry of $\mathbf A$, and
$(\cdot)^{-1}$ denotes the matrix inverse.

\subsection{Problem Formulation}
The BS transmits
$\sum_{k=1}^{K}\sqrt{P_k}\bw_k(\bx_{\rm B})s_k$, where $s_k$
is the unit-power information symbol and $P_k\ge0$ is the power allocated to
user $k$. Define
$\mathbf P=[P_1,\ldots,P_K]^T$ and
$\mathcal P_{\rm B}\triangleq\{\mathbf P\succeq\mathbf0:
\mathbf1^T\mathbf P\le P_{\rm B}\}$, where $P_{\rm B}$ denotes the total
BS transmit-power budget. The jammer sends a unit-power common Gaussian
interference signal using the beamforming vector $\bq\in\mathbb C^M$, where
$\|\bq\|^2\le P_{\rm J}$ and $P_{\rm J}$ is its power budget. Let
$\sigma_k^2$ denote the receiver-noise power at user $k$. Since ZF eliminates
inter-user interference, the achievable rate of user $k$ is
\begin{align}
R_k(\bx_{\rm B},\bx_{\rm J},P_k,\bq)
=\log_2\!\left(1+\frac{P_ka_k(\bx_{\rm B})}
{\sigma_k^2+|\bg_k^H(\bx_{\rm J})\bq|^2}\right).
\label{eq:rate}
\end{align}
Let $\mathcal R(\bx_{\rm B},\bx_{\rm J},\mathbf P,\bq)$ denote the fairness
payoff, i.e., the minimum of
$R_k(\bx_{\rm B},\bx_{\rm J},P_k,\bq)$ over $k\in\cK$.
From the BS side, it aims to jointly optimize its antenna positions
$\bx_{\rm B}$ and user powers $\mathbf P$ to maximize
$\mathcal R(\bx_{\rm B},\bx_{\rm J},\mathbf P,\bq)$. In contrast, the jammer
aims to jointly optimize its antenna positions $\bx_{\rm J}$ and jamming beam
$\bq$ to minimize the same payoff. Therefore, their interaction constitutes
a zero-sum game.

To specify the decision order, we adopt a sequential response model: the BS
first selects $\bx_{\rm B}$, the smart jammer then chooses $(\bx_{\rm J},\bq)$
after observing it, and the BS finally updates $\mathbf P$ according to the
resulting interference. Thus, $\bx_{\rm B}$ must withstand all feasible jammer
actions, whereas $\mathbf P$ can adapt to each realized action and is placed
in the innermost maximization.
Accordingly, the considered zero-sum game is formulated as
\begin{flalign}
&\text{(P1)}\quad
\max_{\bx_{\rm B}\in\cX_{\rm B}}
\min_{\substack{\bx_{\rm J}\in\cX_{\rm J},\,
\|\bq\|^2\le P_{\rm J}}}
\max_{\mathbf P\in\mathcal P_{\rm B}}
\mathcal R(\bx_{\rm B},\bx_{\rm J},\mathbf P,\bq).
&&\label{eq:three-stage}
\end{flalign}
In (P1), $\bx_{\rm B}$ is common to all jammer actions, whereas
$\mathbf P$ is optimized for each particular $(\bx_{\rm J},\bq)$.
Problem (P1) is nonconvex because its three optimization stages are coupled
through the position-dependent channels.

\section{Position-Only Reformulation for General $K$}
\subsection{Elimination of Digital Variables}
We first eliminate the BS power allocation and jamming beam for any given
position pair. The following lemma gives the optimal power allocation for fixed
positions and jamming beam.

\begin{lemma}\label{lem:power}
For notational convenience, define
\begin{align}
c_k(\bx_{\rm B},\bx_{\rm J},\bq)
\triangleq\frac{\sigma_k^2+|\bg_k^H(\bx_{\rm J})\bq|^2}
{a_k(\bx_{\rm B})}.                                         \label{eq:ck}
\end{align}
For fixed $\bx_{\rm B}$, $\bx_{\rm J}$, and $\bq$, the power-allocation
problem is equivalently written as
\begin{align}
\max_{\gamma,\{P_k\}}\ &\gamma
\quad\text{s.t.}\
P_k\ge\gamma c_k(\bx_{\rm B},\bx_{\rm J},\bq),\ \forall k,\
\sum_{k=1}^{K}P_k\le P_{\rm B}.                              \label{eq:power-epi}
\end{align}
Here, $\gamma$ denotes the common signal-to-interference-plus-noise ratio
(SINR). The optimal powers and common SINR are
\begin{align}
P_k^\star(\bx_{\rm B},\bx_{\rm J},\bq)
&=\frac{P_{\rm B}c_k(\bx_{\rm B},\bx_{\rm J},\bq)}
{\sum_{i=1}^{K}c_i(\bx_{\rm B},\bx_{\rm J},\bq)},\nonumber\\
\gamma^\star(\bx_{\rm B},\bx_{\rm J},\bq)
&=\frac{P_{\rm B}}
{\sum_{i=1}^{K}c_i(\bx_{\rm B},\bx_{\rm J},\bq)}.             \label{eq:pstar}
\end{align}
\end{lemma}
\begin{IEEEproof}
Since the logarithm is monotonically increasing, maximizing the minimum rate
is equivalent to maximizing the minimum SINR. A target $\gamma$ requires
$P_ka_k(\bx_{\rm B})/[\sigma_k^2+|\bg_k^H(\bx_{\rm J})\bq|^2]\ge\gamma$,
which gives \eqref{eq:power-epi}. Summing the $K$ constraints yields
$\gamma\sum_kc_k(\bx_{\rm B},\bx_{\rm J},\bq)\le P_{\rm B}$.
Conversely, $P_k=\gamma c_k(\bx_{\rm B},\bx_{\rm J},\bq)$ satisfies all
SINR constraints. Hence, the largest feasible $\gamma$ uses the full BS power
and equalizes the user SINRs, yielding \eqref{eq:pstar}.
\end{IEEEproof}

Since
\begin{align}
\sum_{k=1}^{K}c_k(\bx_{\rm B},\bx_{\rm J},\bq)
=D_0(\bx_{\rm B})+\bq^H\bC(\bx_{\rm B},\bx_{\rm J})\bq,      \label{eq:sum-c}
\end{align}
with $D_0(\bx_{\rm B})\triangleq\sum_{k=1}^{K}\frac{\sigma_k^2}{a_k(\bx_{\rm B})}$ and $\bC(\bx_{\rm B},\bx_{\rm J})\triangleq
\sum_{k=1}^{K}\frac{\bg_k(\bx_{\rm J})\bg_k^H(\bx_{\rm J})}
{a_k(\bx_{\rm B})}$,
the optimized minimum rate based on Lemma~\ref{lem:power} and \eqref{eq:sum-c} can be expressed as
\begin{align}
R_{\min}^{\star}(\bx_{\rm B},\bx_{\rm J},\bq)
=\log_2\!\left(1+\frac{P_{\rm B}}
{D_0(\bx_{\rm B})+\bq^H\bC(\bx_{\rm B},\bx_{\rm J})\bq}\right).
                                                                    \label{eq:rmin-response}
\end{align}
Thus, after observing $\bx_{\rm B}$, the jammer minimizes this power-response
rate by maximizing the denominator of \eqref{eq:rmin-response}.

\begin{proposition}\label{prop:reform}
For any feasible position pair, the optimal jamming beam is
\begin{align}
\bq^\star(\bx_{\rm B},\bx_{\rm J})=\sqrt{P_{\rm J}}\,
\bu_{\max}(\bC(\bx_{\rm B},\bx_{\rm J})),                    \label{eq:qstar}
\end{align}
where $\bu_{\max}(\bC(\bx_{\rm B},\bx_{\rm J}))$ is a principal unit
eigenvector of $\bC(\bx_{\rm B},\bx_{\rm J})$. Moreover, (P1) is equivalent
to
\begin{flalign}
&\text{(P2)}\quad
\min_{\bx_{\rm B}\in\cX_{\rm B}}
\max_{\bx_{\rm J}\in\cX_{\rm J}}
\Phi(\bx_{\rm B},\bx_{\rm J}), &&\notag
\end{flalign}
where $\Phi(\bx_{\rm B},\bx_{\rm J})\triangleq D_0(\bx_{\rm B})+P_{\rm J}
\lambda_{\max}(\bC(\bx_{\rm B},\bx_{\rm J}))$.
\end{proposition}
\begin{IEEEproof}
For fixed positions, $\bC(\bx_{\rm B},\bx_{\rm J})\succeq\mathbf 0$.
Hence, the quadratic form $\bq^H\bC(\bx_{\rm B},\bx_{\rm J})\bq$
is maximized on the boundary $\|\bq\|^2=P_{\rm J}$, i.e., the jammer
always uses its full power. The Rayleigh quotient then gives
\begin{align}
\max_{\|\bq\|^2\le P_{\rm J}}
\bq^H\bC(\bx_{\rm B},\bx_{\rm J})\bq
=P_{\rm J}\lambda_{\max}\!\left(\bC(\bx_{\rm B},\bx_{\rm J})\right),
\end{align}
which is achieved by \eqref{eq:qstar}. Substituting the optimal jamming beam
into \eqref{eq:rmin-response} yields
\begin{align}
R_{\min}^{\star}(\bx_{\rm B},\bx_{\rm J})
=\log_2\!\left(
1+\frac{P_{\rm B}}{\Phi(\bx_{\rm B},\bx_{\rm J})}
\right).
\end{align}
Since $\log_2(1+P_{\rm B}/x)$ is strictly decreasing for $x>0$, maximizing
the worst-case rate over $\bx_{\rm B}$ is equivalent to minimizing
$\Phi(\bx_{\rm B},\bx_{\rm J})$ in (P2).
\end{IEEEproof}

Lemma~\ref{lem:power} and Proposition~\ref{prop:reform} exactly eliminate the
$K$ BS powers and the complex jamming beam; no relaxation has been introduced.
Only the $N+M$ real position variables remain. Nevertheless, (P2) is a highly
nonconvex zero-sum position game because of the complex-exponential channels,
the inverse Gram matrix, and the maximum-eigenvalue term.

\subsection{Closed-Form Whole-Array Position Gradients}
To solve (P2) through vector updates, we derive the gradients of
$\Phi(\bx_{\rm B},\bx_{\rm J})$ with respect to the complete BS and jammer
position vectors. Define
\begin{align}
\bZ(\bx_{\rm B})&\triangleq\bG^{-1}(\bx_{\rm B}),\nonumber\\
\boldsymbol{\nu}(\bx_{\rm B},\bx_{\rm J})
&\triangleq\bu_{\max}\!\left(\bC(\bx_{\rm B},\bx_{\rm J})\right),
                                                                    \label{eq:z-nu}\\
\bJ(\bx_{\rm J})&\triangleq
[\bg_1(\bx_{\rm J}),\ldots,\bg_K(\bx_{\rm J})],\nonumber\\
\bu&\triangleq[u_1,\ldots,u_K]^T,\quad
\bv\triangleq[v_1,\ldots,v_K]^T,\nonumber\\
\bz(\bx_{\rm B})&\triangleq
[[\bZ(\bx_{\rm B})]_{11},\ldots,[\bZ(\bx_{\rm B})]_{KK}]^T.
                                                        \label{eq:compact-vectors}
\end{align}
Here, $\boldsymbol{\nu}(\bx_{\rm B},\bx_{\rm J})$ is selected as a
unit-norm principal eigenvector. Further define
\begin{align}
\chi_k(\bx_{\rm B},\bx_{\rm J})
&\triangleq\bg_k^H(\bx_{\rm J})
\boldsymbol{\nu}(\bx_{\rm B},\bx_{\rm J}),\nonumber\\
\boldsymbol{\chi}(\bx_{\rm B},\bx_{\rm J})
&\triangleq[\chi_1(\bx_{\rm B},\bx_{\rm J}),\ldots,
\chi_K(\bx_{\rm B},\bx_{\rm J})]^T,\nonumber\\
\boldsymbol{\xi}(\bx_{\rm B},\bx_{\rm J})
&\triangleq\bz(\bx_{\rm B})\odot\bv\odot
\boldsymbol{\chi}^*(\bx_{\rm B},\bx_{\rm J}),\nonumber\\
\omega_k(\bx_{\rm B},\bx_{\rm J})
&\triangleq\sigma_k^2+P_{\rm J}
|\chi_k(\bx_{\rm B},\bx_{\rm J})|^2,\nonumber\\
\boldsymbol{\Omega}(\bx_{\rm B},\bx_{\rm J})
&\triangleq\operatorname{diag}
\bigl(\omega_1(\bx_{\rm B},\bx_{\rm J}),\ldots,
\omega_K(\bx_{\rm B},\bx_{\rm J})\bigr),                    \label{eq:omega}\\
\bT(\bx_{\rm B},\bx_{\rm J})
&\triangleq\bH(\bx_{\rm B})\bZ(\bx_{\rm B})
\boldsymbol{\Omega}(\bx_{\rm B},\bx_{\rm J})\bZ(\bx_{\rm B}).
                                                                  \label{eq:T}
\end{align}

\begin{lemma}\label{lem:gradient}
Suppose that $\bG(\bx_{\rm B})\succ\mathbf0$ and the largest eigenvalue of
$\bC(\bx_{\rm B},\bx_{\rm J})$ is simple. Then, the whole-array gradients are
\begin{align}
\nabla_{\bx_{\rm B}}\Phi(\bx_{\rm B},\bx_{\rm J})
&=2\kappa\Ima\!\left\{
[\bH(\bx_{\rm B})\odot\bT^*(\bx_{\rm B},\bx_{\rm J})]\bu
\right\},                                             \label{eq:closed-grad-b}\\
\nabla_{\bx_{\rm J}}\Phi(\bx_{\rm B},\bx_{\rm J})
&=2\kappa P_{\rm J}\Ima\!\left\{
\operatorname{diag}(\boldsymbol{\nu}(\bx_{\rm B},\bx_{\rm J}))
\bJ^*(\bx_{\rm J})\right.\nonumber\\[-1mm]
&\quad\left.\times
\boldsymbol{\xi}(\bx_{\rm B},\bx_{\rm J})\right\}.
                                                        \label{eq:closed-grad-j}
\end{align}
Here, $\odot$ denotes the Hadamard product, $(\cdot)^*$ denotes elementwise
complex conjugation, and $\Ima\{\cdot\}$ is applied elementwise. Therefore,
\eqref{eq:closed-grad-b} and \eqref{eq:closed-grad-j} directly give
$N$- and $M$-dimensional real gradient vectors, respectively.
\end{lemma}

\begin{IEEEproof}
From \eqref{eq:ak}, we have
$1/a_k(\bx_{\rm B})=[\bZ(\bx_{\rm B})]_{kk}$. Therefore,
\begin{align}
\Phi(\bx_{\rm B},\bx_{\rm J})
&=\sum_{k=1}^{K}\sigma_k^2[\bZ(\bx_{\rm B})]_{kk}
+P_{\rm J}\lambda_{\max}
\bigl(\bC(\bx_{\rm B},\bx_{\rm J})\bigr).              \label{eq:phi-expanded}
\end{align}
For a simple largest eigenvalue and its unit principal eigenvector,
\begin{align}
\mathrm d\lambda_{\max}
\bigl(\bC(\bx_{\rm B},\bx_{\rm J})\bigr)
&=\boldsymbol{\nu}^H(\bx_{\rm B},\bx_{\rm J})
\mathrm d\bC(\bx_{\rm B},\bx_{\rm J})
\boldsymbol{\nu}(\bx_{\rm B},\bx_{\rm J}).              \label{eq:eigen-diff}
\end{align}
Based on \eqref{eq:phi-expanded} and \eqref{eq:eigen-diff}, we next derive
the gradients with respect to $\bx_{\rm B}$ and $\bx_{\rm J}$ separately.

First, vary $\bx_{\rm B}$ while fixing $\bx_{\rm J}$. In this case,
\begin{align}
\mathrm d\bC(\bx_{\rm B},\bx_{\rm J})
=\sum_{k=1}^{K}\mathrm d[\bZ(\bx_{\rm B})]_{kk}
\bg_k(\bx_{\rm J})\bg_k^H(\bx_{\rm J}).                 \label{eq:diff-cb}
\end{align}
Substituting \eqref{eq:diff-cb} into \eqref{eq:eigen-diff} and
differentiating \eqref{eq:phi-expanded} give
\begin{align}
\mathrm d\Phi(\bx_{\rm B},\bx_{\rm J})
=\Tr\!\left(
\boldsymbol{\Omega}(\bx_{\rm B},\bx_{\rm J})
\mathrm d\bZ(\bx_{\rm B})\right),                       \label{eq:diff-z}
\end{align}
where $\boldsymbol{\Omega}(\bx_{\rm B},\bx_{\rm J})$ collects the
coefficients of the diagonal entries of $\mathrm d\bZ(\bx_{\rm B})$.
Moreover,
\begin{align}
\mathrm d\bZ(\bx_{\rm B})
&=-\bZ(\bx_{\rm B})\mathrm d\bG(\bx_{\rm B})\bZ(\bx_{\rm B}),\nonumber\\
\mathrm d\bG(\bx_{\rm B})
&=[\mathrm d\bH(\bx_{\rm B})]^H\bH(\bx_{\rm B})
+\bH^H(\bx_{\rm B})\mathrm d\bH(\bx_{\rm B}),\nonumber\\
\mathrm d\bH(\bx_{\rm B})
&=\mathrm j\kappa\operatorname{diag}(\mathrm d\bx_{\rm B})
\bH(\bx_{\rm B})\operatorname{diag}(\bu).               \label{eq:diff-h-matrix}
\end{align}
Since $\bZ(\bx_{\rm B})$ and
$\boldsymbol{\Omega}(\bx_{\rm B},\bx_{\rm J})$ are Hermitian,
$\bZ(\bx_{\rm B})\boldsymbol{\Omega}(\bx_{\rm B},\bx_{\rm J})
\bZ(\bx_{\rm B})$ is also Hermitian. Using \eqref{eq:T}, the cyclic trace
property, and \eqref{eq:diff-h-matrix}, we obtain
\begin{align}
&\mathrm d\Phi(\bx_{\rm B},\bx_{\rm J})
=-2\Rea\!\left\{\Tr\!\left(
\bT^H(\bx_{\rm B},\bx_{\rm J})\mathrm d\bH(\bx_{\rm B})
\right)\right\}\nonumber\\
=&\left[2\kappa\Ima\!\left\{
[\bH(\bx_{\rm B})\odot\bT^*(\bx_{\rm B},\bx_{\rm J})]\bu
\right\}\right]^T\mathrm d\bx_{\rm B}.                 \label{eq:diff-b-final}
\end{align}
This proves \eqref{eq:closed-grad-b}.

Next, vary $\bx_{\rm J}$ while fixing $\bx_{\rm B}$. We have
\begin{align}
\mathrm d\bC(\bx_{\rm B},\bx_{\rm J})
=\sum_{k=1}^{K}[\bZ(\bx_{\rm B})]_{kk}
\bigl(&[\mathrm d\bg_k(\bx_{\rm J})]\bg_k^H(\bx_{\rm J})\nonumber\\[-1mm]
&+\bg_k(\bx_{\rm J})[\mathrm d\bg_k(\bx_{\rm J})]^H\bigr),    \label{eq:diff-cj}
\end{align}
and
\begin{align}
\mathrm d\bg_k(\bx_{\rm J})
=\mathrm j\kappa v_k\operatorname{diag}(\bg_k(\bx_{\rm J}))
\mathrm d\bx_{\rm J}.                                      \label{eq:diff-g}
\end{align}
Substituting \eqref{eq:diff-cj} and \eqref{eq:diff-g} into
\eqref{eq:eigen-diff} and using \eqref{eq:phi-expanded} yield
\begin{align}
\mathrm d\Phi(\bx_{\rm B},\bx_{\rm J})
&=2P_{\rm J}\sum_{k=1}^{K}[\bZ(\bx_{\rm B})]_{kk}
\Rea\!\left\{\chi_k^*(\bx_{\rm B},\bx_{\rm J})\right.\nonumber\\[-1mm]
&\quad\left.\times
[\mathrm d\bg_k(\bx_{\rm J})]^H
\boldsymbol{\nu}(\bx_{\rm B},\bx_{\rm J})\right\}\nonumber\\
&=2\kappa P_{\rm J}\Bigl[\Ima\!\left\{
\operatorname{diag}(\boldsymbol{\nu}(\bx_{\rm B},\bx_{\rm J}))
\bJ^*(\bx_{\rm J})\right.\nonumber\\[-1mm]
&\quad\left.\times
\boldsymbol{\xi}(\bx_{\rm B},\bx_{\rm J})\right\}\Bigr]^T
\mathrm d\bx_{\rm J}.                                  \label{eq:diff-j-final}
\end{align}
This proves \eqref{eq:closed-grad-j}.
\end{IEEEproof}

\begin{figure*}[!t]
\vspace{-8pt}
\centering
\includegraphics[width=0.92\textwidth]{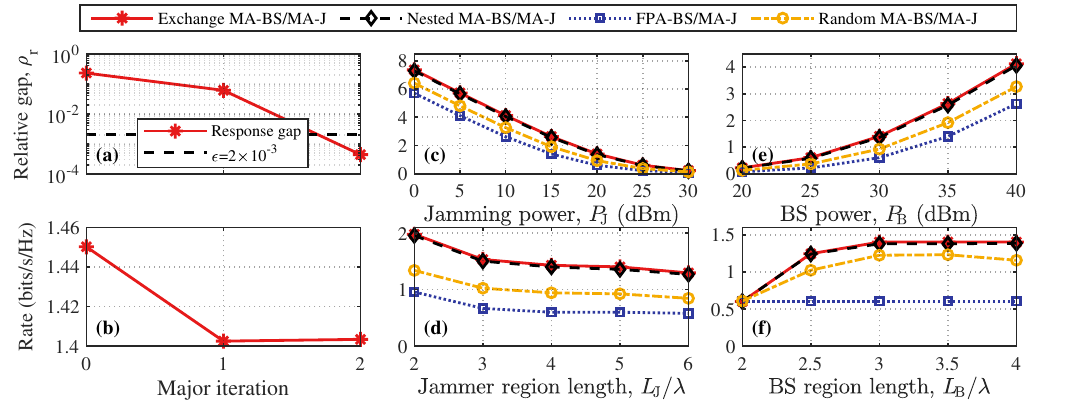}
\caption{General-$K$ results: (a)--(b) response gap and response rate versus
major iteration; (c)--(d) independently verified minimum rate versus the
jamming power and jammer movement-region length; (e)--(f) the corresponding
rate versus the BS power and BS movement-region length.}
\label{fig:results}
\end{figure*}

\subsection{Exchange-Based Position Optimization}
Based on the closed-form gradients in Lemma~\ref{lem:gradient}, we next
develop an exchange-based method for solving (P2)
\cite{ZhangExchange2010,ZengCCG2013}. Specifically, by introducing
an epigraph variable $\tau$, (P2) is equivalently rewritten as
\begin{equation}
\text{(P3)}\;\min_{\bx_{\rm B}\in\cX_{\rm B},\,\tau}\tau\;
\mathrm{s.t.}\;\Phi(\bx_{\rm B},\bx_{\rm J})\le\tau,\;
\forall\bx_{\rm J}\in\cX_{\rm J}.                       \label{eq:sip}
\end{equation}
Problem (P3) is semi-infinite because each feasible jammer deployment gives one
constraint. A nested method repeatedly solves the jammer problem at every BS
iterate and line-search point, which requires many jammer searches
\cite{ZhangExchange2010,ZengCCG2013}. In
contrast, an exchange scheme reuses previously detected attacks during the BS
updates and invokes a complete jammer search only in a separation step.
Motivated by this computational advantage, we adopt an exchange method for
(P3) \cite{ZhangExchange2010,ZengCCG2013}. Specifically, its finite master
step obtains a BS candidate against the stored attacks, and its separation
step verifies this candidate over the complete jammer feasible set. Only a
sufficiently violating response is added to the attack set.
The finite master is
optimistic because it protects the BS only against detected attacks. Separation
searches for an omitted constraint at the returned candidate; adding a
violating response forces the next master step to correct the exposed
robustness loss.

\emph{1) Finite master problem:}
At major iteration $r$, let
$\mathcal S_{\rm J}^{(r)}=\{\bx_{\rm J}^{[1]},\ldots,
\bx_{\rm J}^{[Q_r]}\}\subseteq\cX_{\rm J}$ denote the attack set. Retaining
only its $Q_r$ constraints gives
\begin{flalign}
&\text{(P4-$r$)}\quad
\min_{\bx_{\rm B}\in\cX_{\rm B}}
\max_{1\le\ell\le Q_r}
\Phi(\bx_{\rm B},\bx_{\rm J}^{[\ell]}). &&               \label{eq:finite-master}
\end{flalign}
Since the finite maximum is nonsmooth, we use its log-sum-exp approximation
\begin{align}
F_{\mu,r}(\bx_{\rm B})
\triangleq\mu\log\sum_{\ell=1}^{Q_r}
\exp\!\left(\Phi(\bx_{\rm B},\bx_{\rm J}^{[\ell]})/\mu\right),
                                                                  \label{eq:lse}
\end{align}
where $\mu>0$ controls the smoothness and approximation accuracy. Let
$\Phi_{\ell}(\bx_{\rm B})\triangleq
\Phi(\bx_{\rm B},\bx_{\rm J}^{[\ell]})$. The softmax weights and the
corresponding weighted matrix are
\begin{align}
\pi_{\ell}(\bx_{\rm B})
&\triangleq\frac{\exp(\Phi_{\ell}(\bx_{\rm B})/\mu)}
{\sum_{i=1}^{Q_r}\exp(\Phi_i(\bx_{\rm B})/\mu)},\nonumber\\
\overline{\bT}_{\mu,r}(\bx_{\rm B})
&\triangleq\sum_{\ell=1}^{Q_r}\pi_{\ell}(\bx_{\rm B})
\bT(\bx_{\rm B},\bx_{\rm J}^{[\ell]}).                    \label{eq:weighted-T}
\end{align}
As $\mu$ decreases, the approximation becomes closer to the finite maximum;
a larger $\mu$, in contrast, produces a smoother objective. Moreover, the
nonnegative weights sum to one and emphasize the more harmful stored
deployments. Then, under the conditions of Lemma~\ref{lem:gradient}, the chain
rule gives
\begin{align}
\nabla_{\bx_{\rm B}}F_{\mu,r}(\bx_{\rm B})
&=\sum_{\ell=1}^{Q_r}\pi_{\ell}(\bx_{\rm B})
\nabla_{\bx_{\rm B}}\Phi_{\ell}(\bx_{\rm B})\nonumber\\
&=2\kappa\Ima\!\left\{
[\bH(\bx_{\rm B})\odot
\overline{\bT}_{\mu,r}^{*}(\bx_{\rm B})]\bu
\right\}.                                                   \label{eq:master-grad}
\end{align}

To minimize $F_{\mu,r}$, we employ projected-gradient descent. Specifically,
at inner iteration $n$, let $\mathbf d_{\rm B}^{(n)}\triangleq-
\nabla_{\bx_{\rm B}}F_{\mu,r}(\bx_{\rm B}^{(n)})$ and update
\begin{align}
\bx_{\rm B}^{(n+1)}
=\arg\min_{\bx_{\rm B}\in\cX_{\rm B}}\frac{1}{2}
\left\|\bx_{\rm B}-\left(\bx_{\rm B}^{(n)}+
\eta_{\rm B}^{(n)}\mathbf d_{\rm B}^{(n)}\right)\right\|_2^2.
                                                               \label{eq:master-update}
\end{align}
The projection directly enforces the ordered position constraints, which are
linear after rewriting \eqref{eq:set}. Since its objective is strictly convex
and $\cX_{\rm B}$ is closed and convex, the projected point is unique. The
trial step $\eta_{\rm B}^{(n)}$ is repeatedly reduced by backtracking until a
sufficient decrease of $F_{\mu,r}$ is obtained.

Since the master problem is nonconvex, we use $S_{\rm B}$ starts. Let
$\widetilde\bx_{\rm B,s}^{(r)}\in\cX_{\rm B}$ be a random feasible point and
$\bx_{\rm B,s}^{(r,{\rm end})}$ be the terminal point from start $s$.
Moreover, let $\mathcal I_{\rm B}^{(r)}$ denote the set of the $S_{\rm B}$
initial points supplied to the finite-master solver at major iteration $r$.
Then,
\begin{align*}
\mathcal I_{\rm B}^{(0)}
&=\{\widetilde\bx_{\rm B,s}^{(0)}\}_{s=1}^{S_{\rm B}},\quad
\mathcal I_{\rm B}^{(r)}
=\{\bx_{\rm B}^{(r-1)}\}\cup
\{\widetilde\bx_{\rm B,s}^{(r)}\}_{s=2}^{S_{\rm B}},\ r\ge1,\\
\bx_{\rm B}^{(r)}
&\in\arg\min_{1\le s\le S_{\rm B}}
F_{\mu,r}(\bx_{\rm B,s}^{(r,{\rm end})}).
\end{align*}
Thus, all points in $\mathcal I_{\rm B}^{(0)}$ are random. For $r\ge1$, start
1 is initialized as $\widetilde\bx_{\rm B,1}^{(r)}=\bx_{\rm B}^{(r-1)}$,
whereas starts $s=2,\ldots,S_{\rm B}$ are the new random points
$\widetilde\bx_{\rm B,s}^{(r)}$.

\emph{2) Jammer separation:}
The finite master only covers the stored attacks. Therefore, for the obtained
$\bx_{\rm B}^{(r)}$, the separation step solves
\begin{align}
\max_{\bx_{\rm J}\in\cX_{\rm J}}
\Phi(\bx_{\rm B}^{(r)},\bx_{\rm J}).                       \label{eq:separation}
\end{align}
To this end, we use projected-gradient ascent. At inner iteration
$n$, let $\mathbf d_{\rm J}^{(n)}\triangleq\nabla_{\bx_{\rm J}}
\Phi(\bx_{\rm B}^{(r)},\bx_{\rm J}^{(n)})$ and update
\begin{align}
\bx_{\rm J}^{(n+1)}
=\arg\min_{\bx_{\rm J}\in\cX_{\rm J}}\frac{1}{2}
\left\|\bx_{\rm J}-\left(\bx_{\rm J}^{(n)}+
\eta_{\rm J}^{(n)}\mathbf d_{\rm J}^{(n)}\right)\right\|_2^2,
                                                               \label{eq:separation-update}
\end{align}
where backtracking now requires a sufficient increase of $\Phi$.

Since the separation problem is nonconvex, we use $S_{\rm J}$ starts. Let
$\bx_{\rm J,old}^{(r)}$ be the strongest deployment in
$\mathcal S_{\rm J}^{(r)}$ under the current $\bx_{\rm B}^{(r)}$. Denote
$\widetilde\bx_{\rm J,s}^{(r)}\in\cX_{\rm J}$ as a new random feasible point,
and $\bx_{\rm J,s}^{(r,{\rm end})}$ as the terminal point from start $s$.
Moreover, let $\mathcal I_{\rm J}^{(r)}$ denote the set of the $S_{\rm J}$
initial points supplied to the separation solver at major iteration $r$. Then,
\begin{align*}
\mathcal I_{\rm J}^{(r)}
&=\{\bx_{\rm J,old}^{(r)}\}\cup
\{\widetilde\bx_{\rm J,s}^{(r)}\}_{s=2}^{S_{\rm J}},\\
\widehat\bx_{\rm J}^{(r)}
&\in\arg\max_{1\le s\le S_{\rm J}}
\Phi(\bx_{\rm B}^{(r)},\bx_{\rm J,s}^{(r,{\rm end})}).
\end{align*}
Thus, start 1 in $\mathcal I_{\rm J}^{(r)}$ is initialized as
$\widetilde\bx_{\rm J,1}^{(r)}=\bx_{\rm J,old}^{(r)}$, whereas starts
$s=2,\ldots,S_{\rm J}$ are the new random points
$\widetilde\bx_{\rm J,s}^{(r)}$. Since backtracking accepts only a sufficient
increase, the terminal point from start 1 is no weaker than
$\bx_{\rm J,old}^{(r)}$; hence, it is sufficient to select the best terminal
point.

Accordingly, let
\begin{align}
V_r&\triangleq\max_{\bx_{\rm J}\in\mathcal S_{\rm J}^{(r)}}
\Phi(\bx_{\rm B}^{(r)},\bx_{\rm J}),\nonumber\\
\Delta_r&\triangleq
\Phi(\bx_{\rm B}^{(r)},\widehat\bx_{\rm J}^{(r)})-V_r,
                                                        \label{eq:violation}\\
\rho_r&\triangleq\Delta_r/V_r.                    \label{eq:relative-gap}
\end{align}
Here, $\Delta_r$ and $\rho_r$ are the absolute and relative response gaps.
Since $\Phi>0$, we have $V_r>0$. Moreover, the warm start and
sufficient-increase backtracking ensure $\Delta_r\ge0$, and hence
$\rho_r\ge0$. We terminate when $\rho_r\le\epsilon$; otherwise,
\begin{align}
\mathcal S_{\rm J}^{(r+1)}
=\mathcal S_{\rm J}^{(r)}\cup
\{\widehat\bx_{\rm J}^{(r)}\}.                            \label{eq:set-update}
\end{align}
Thus, the next finite master problem explicitly accounts for the newly
detected attack.

The tolerance $\epsilon>0$ controls accuracy versus complexity: a smaller
value is stricter but may require more major iterations. The adopted
$\epsilon=2\times10^{-3}$ limits the newly found relative improvement to
$0.2\%$. Equivalently, the stopping test $\rho_r\le\epsilon$ requires
$\Phi(\bx_{\rm B}^{(r)},\widehat\bx_{\rm J}^{(r)})\le
(1+\epsilon)V_r=1.002V_r$.

\par\vspace{8pt}\noindent\begin{minipage}{\columnwidth}
\refstepcounter{algorithm}\label{alg:exchange}
\hrule
\noindent\textbf{Algorithm \thealgorithm: Exchange method for (P3)}
\hrule
\begin{tabular}{@{}r@{\hspace{0.5em}}p{\dimexpr\columnwidth-2.0em\relax}@{}}
&\textbf{Input:} $\mathcal S_{\rm J}^{(0)}$, $\mu$, $\epsilon$, $S_{\rm B}$,
$S_{\rm J}$, and iteration limits.\\[-2pt]
1:&Set $r=0$.\\[-2pt]
2:&Form $\mathcal I_{\rm B}^{(r)}$ and solve the finite master to obtain
$\bx_{\rm B}^{(r)}$.\\[-2pt]
3:&Form $\mathcal I_{\rm J}^{(r)}$ and run separation to obtain
$\widehat\bx_{\rm J}^{(r)}$ and $\rho_r$.\\[-2pt]
4:&If $\rho_r\le\epsilon$, terminate. Otherwise, update the attack set by
\eqref{eq:set-update}, set $r\leftarrow r+1$, and return to Step 2.\\[-2pt]
&\textbf{Output:} $\bx_{\rm B}^{(r)}$ and $\widehat\bx_{\rm J}^{(r)}$.
\end{tabular}
\hrule
\end{minipage}\par\vspace{7pt}

\emph{3) Complexity and nested benchmark:}
Finally, let $I_{\rm M}$ and $I_{\rm J}$ be the average evaluation numbers per
master and jammer start, respectively, $C_{\rm g}$ be the objective-gradient
cost, $I_{\rm N}$ be the number of nested BS updates, and $B_{\rm N}$ be the
average jammer searches per update. For $R$ major iterations,
let $\mathcal C_{\rm ex}$ and $\mathcal C_{\rm ne}$ denote the complexities of
the proposed exchange method and nested benchmark, respectively:
\begin{align}
\mathcal C_{\rm ex}
&=\mathcal O\!\left(\sum_{r=0}^{R-1}
[S_{\rm B}I_{\rm M}Q_r+S_{\rm J}I_{\rm J}]C_{\rm g}\right),
                                                        \label{eq:complexity}\\
\mathcal C_{\rm ne}
&=\mathcal O\!\left(
S_{\rm B}I_{\rm N}B_{\rm N}S_{\rm J}I_{\rm J}C_{\rm g}\right).
                                                 \label{eq:nested-complexity}
\end{align}
Thus, exchange performs a complete jammer search only once per major iteration
and reuses the $Q_r$ stored attacks, whereas the nested method invokes an
$S_{\rm J}$-start search at each BS iterate and backtracking point.

\section{Numerical Results}
Unless otherwise specified, we set $N=M=5$, $K=4$, a $3$-GHz carrier,
$d_{\rm B}=d_{\rm J}=\lambda/2$, $P_{\rm B}=30$ dBm, and
$\sigma_k^2=-80$ dBm. The channel gains are
$\beta_k=\beta_0(D_{{\rm B},k}/D_0)^{-\eta_{\rm B}}$ and
$\alpha_k=\beta_0(D_{{\rm J},k}/D_0)^{-\eta_{\rm J}}$, with $D_0=1$ m,
$10\log_{10}\beta_0=-30$ dB, and $\eta_{\rm B}=\eta_{\rm J}=2.2$. We set
$\bD_{\rm B}=[50,55,60,65]^T$ m and
$\bD_{\rm J}=[30,35,40,45]^T$ m,
$\bu=[-0.45,-0.18,0.09,0.36]^T$ and
$\bv=[-0.52,-0.17,0.19,0.47]^T$, with default
$L_{\rm B}=L_{\rm J}=5\lambda$ and $P_{\rm J}=20$ dBm. The initial attack set
contains the FPA, a spread array, and one random deployment. We use
$\mu=10^{-3}$, $\epsilon=2\times10^{-3}$, at most nine major iterations,
$S_{\rm B}=4$, $S_{\rm J}=5$, 45 master steps, and 60 separation steps. The
nested benchmark uses the same starts and at most 24 outer steps. Both methods
use Armijo backtracking, and each BS deployment is verified using the FPA/spread
starts, 12 random jammer starts, and 100 ascent steps.

We compare \emph{Exchange MA-BS/MA-J} with \emph{Nested MA-BS/MA-J},
\emph{FPA-BS/MA-J}, and \emph{Random MA-BS/MA-J}. The proposed exchange scheme
follows Algorithm~\ref{alg:exchange}. The nested scheme uses the same multistart
updates but repeats a complete jammer search at every BS iterate and
backtracking point. FPA-BS/MA-J fixes a half-wavelength BS ULA. Random
MA-BS/MA-J averages 20 feasible random BS deployments without BS position
optimization. In both baselines, the jammer positions are optimized. All
schemes use \eqref{eq:pstar}--\eqref{eq:qstar} and the same verification.

Figs.~\ref{fig:results}(a) and (b) show that, as major iteration $r$ increases
from $0$ to $2$, $\rho_r$ decreases from $0.227$ through $0.0604$ to
$4.34\times10^{-4}$ and falls below $\epsilon=2\times10^{-3}$, while the
response rate stabilizes near $1.403$ bits/s/Hz. At the default setting,
exchange and nested respectively use $6233$ and $51835$ objective-gradient
evaluations, corresponding to an $8.3$-fold reduction.

Figs.~\ref{fig:results}(c) and (d) show that all rates decrease with the
jammer-side resources. At $P_{\rm J}=20$ dBm, exchange, nested, random MA-BS,
and FPA-BS achieve $1.403$, $1.360$, $0.926$, and $0.601$ bits/s/Hz,
respectively. Increasing $L_{\rm J}$ from $2\lambda$ to $6\lambda$ reduces
their exchange and random-BS rates from $1.980$ and $1.341$ to $1.297$ and
$0.848$ bits/s/Hz, respectively. For $L_{\rm J,1}\le L_{\rm J,2}$, the
jammer feasible sets are nested; hence, its maximized $\Phi$ cannot decrease,
while the enlarged region provides more phase-alignment freedom. Since the
rate is $\log_2(1+P_{\rm B}/\Phi)$, its worst-case value decreases.

Figs.~\ref{fig:results}(e) and (f) show that increasing $P_{\rm B}$ from $20$
to $40$ dBm raises the exchange rate from $0.220$ to $4.125$ bits/s/Hz. At
$L_{\rm B}=3\lambda$, exchange achieves $1.404$ bits/s/Hz, compared with
$1.226$ and $0.607$ bits/s/Hz for random MA-BS and FPA-BS, respectively. Thus,
optimized BS movement provides a clear gain against the optimized MA jammer.

\section{Conclusion}
This paper studied fair downlink with linear MAs at the BS and jammer.
Closed-form digital strategies reduced the game to a position-only min--max
problem. An exchange method then used whole-array gradients and ordered-set
projections to achieve competitive fairness with substantially fewer
objective-gradient evaluations.

\bibliographystyle{IEEEtran}
\bibliography{references}

\end{document}